\documentstyle[]{l-aa}

\newcommand{\msol}{{M$_{\odot}$}}
\newcommand{\msolyr}{{M$_{\odot}$}yr$^{-1}$ }
\newcommand{\lsol}{{L$_{\odot}$}}

\newcommand{\kks}{K km s$^{-1}$}
\newcommand{\ks}{km s$^{-1}$}

\include{psfig}

\begin{document}

\thesaurus{06(08.03.4; 08.09.2: IRC~+10~216; 08.13.2; 08.16.4; 13.19.5)}

\title{New peculiar CO data of the shell around IRC~+10~216 
}

\author{M.~A.~T.~Groenewegen \inst{1}
\and H.-G.~Ludwig \inst{2,1}
}

\offprints{Martin Groenewegen (groen@mpa-garching.mpg.de)}

\institute{Max-Planck-Institut f\"ur Astrophysik, 
Karl-Schwarzschild-Stra{\ss}e 1, D-85740 Garching, Germany
\and
Astronomical Observatory, Niels Bohr Institute, 
Juliane Maries Vej 30, DK-2100 Copenhagen, Denmark
}

\date{received,  accepted}

\maketitle
\markboth{Groenewegen \& Ludwig: New peculiar CO data of the shell 
around IRC~+10~216}{}

\begin{abstract}
A CO(1-0) on-source spectrum of the well-known carbon star IRC~+10~216
taken with the IRAM 30m telescope in June 1996 shows excess emission
between --18.3 and --14.3 \ks\ at the red wing of the underlying
profile. The excess emission is confirmed in January 1997 but is
absent in April 1997 and June 1998 IRAM spectra. Such a transient
feature has not been seen before in this star or any other AGB star.
In April 1997 we mapped the circumstellar shell out to
110\arcsec. Both the J = 1-0 and J = 2-1 spectra show ``spikes'' or
components which vary in strength with position in the envelope. One
of these components corresponds to the velocity interval mentioned
above. An immediate conclusion is that the circumstellar shell is not
spherically symmetric, contrary to what was believed based on lower
spectral resolution data. We are probably seeing emission from a
complex geometrical structure. Neither a disk structure nor a
double-wind structure seem to be able to explain the observations.
The on-source transient behaviour of the red excess emission can
reasonably well be explained by a single large ($\sim 5 \times
10^{13}$ cm) blob, that expands due to internal motion.

\end{abstract}

\section{Introduction}

IRC~+10~216 (= AFGL~1381 = CW~Leo) is often considered the prototype
infrared carbon star. It is a carbon Mira with a period of 649 days
(Le Bertre 1992) and of considerable apparent brightness. It is
luminous, about 10~000 \lsol\ following the period-luminosity relation
of Groenewegen \& Whitelock (1996), and at a distance of about 135 pc
(e.g. Le Bertre 1997, Groenewegen et al. 1998).  It is a well-known
calibration object, in particular for molecular line emission
observations (e.g. Mauersberger et al. 1989).

Here we report on excess emission seen in a CO(1-0) spectrum taken on
25 June 1996 with the IRAM telescope, and confirmed by later
observations. We present a map of the circumstellar shell at high
velocity resolution. We compare the results to previous CO observations.

\begin{figure}
\centerline{\psfig{figure=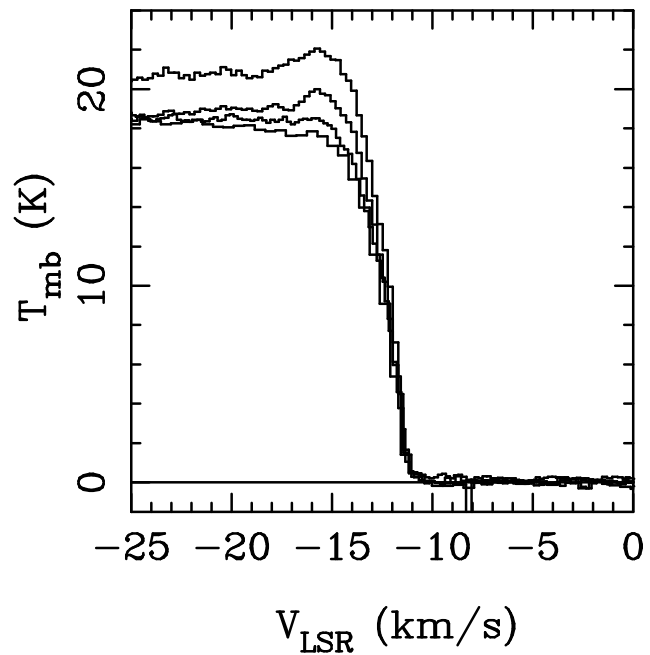,width=8.99cm,angle=-90}}
\caption[]{Red part of the CO J = 1-0 spectrum. From top to bottom
June 1996, January 1997 (both at 100 KHz resolution), April 1997
spectrum at 78 KHz resolution, June 1998 (at 200 KHz resolution). The
rms noise is better than 0.17 K. The June 1998 spectrum has been
multiplied with a factor of 0.91 for clarity. 
The others are calibrated spectra as observed.
}
\end{figure}

\section{The observations}

The IRAM observations were obtained on June 25, 1996 (observer HGL),
January 15-16, 1997 (observer MG), April 3-4, 1997 (observer MG) and
June 3-4, 1998 (observer MG).  On the first two occasions the
instrumental set-up was identical.  Two 1.3mm SIS receivers and two
3.0mm SIS receivers (measuring the different polarizations) were used
simultaneously and tuned to the CO(1-0) and CO(2-1) lines. The 100 KHz
backend was connected to one of the 3mm receivers, resulting in a
velocity separation per channel of 0.26 \ks. The 1MHz backend was
split in two and connected to the two 1.3mm receivers, resulting in a
velocity separation of 1.3 \ks. The autocorrelator was split into two
and connected to the two 1.3mm receivers resulting in a velocity
separation of 0.81 \ks. In April 1997 the set-up was as follows: the
two 3mm receivers were tuned to the CO (1-0) and the HCN(1-0) line,
while the 1.3mm receiver was tuned to the CO (2-1) line. The 1MHz
backend was used to observe the HCN line (channel separation of 3.38
\ks). The autocorrelator was split to observe the CO (1-0) line at a
resolution of 78 KHz (0.20 \ks), and the J = 2-1 line at 312 KHz
resolution (or 0.41 \ks). In June 1998 the set-up was as on the first
two epochs using the 1MHz backend and the autocorrelator.

IRC~+10~216 was observed on two consecutive occasions on June 25, 1996
and seven times between January 15-16, 1997 in the course of another
project as calibration observations on both occasions.  In April 1997
we mapped the star out to a radial off-set of about 110\arcsec\ in the
CO (1-0), CO (2-1) and HCN (1-0) lines. In June 1998 the star was
observed once for calibration purposes on another project.

The high-resolution spectra of the J = 1-0 line are shown in Fig.~1.
The blue part is missing in the June 1996 and January 1997
observations due to the limited number of channels of the backend and
because we did not center it on the line.  The June 1996 profile shows
excess emission between --18.3 and --14.3 \ks.  A technical defect can
be ruled out as the emission is visible in all six spectra taken with
different receivers and different backends.  In January 1997 we also
checked that the emission was still present if we disconnected one of
the 3mm and 1.3mm receivers, and the corresponding backend. An
interference is very unlikely as this affects mostly one channel while
the excess emission is several \ks\ wide.  In June 1996, the peak
temperature\footnote{The temperature scale in this paper is mean-beam
brightness temperature.}  of the excess emission is 1.2 K with respect
to the underlying plateau and has an integrated intensity of 2.4
\kks. The integrated intensity of the entire CO (1-0) lines as
measured with the autocorrelator is 544 \kks. In January 1997, the
total integrated intensity was 505 \kks, the excess emission in the
velocity interval between --18.4 and --14.3 \ks\ was 1.7 \kks, and the
peak was 1.6 K higher than the line center. It is also clear that even
at bluer velocities the intensity is less, rising towards the
red. This is particularly clear when compared to the April 1997
observation when the 1-0 profile was flat-topped again and the excess
emission has disappeared. In June 1998 this corresponds even to a
``lack'' of emission with respect to the flat-topped profile.

\begin{figure*}
\centerline{\psfig{figure=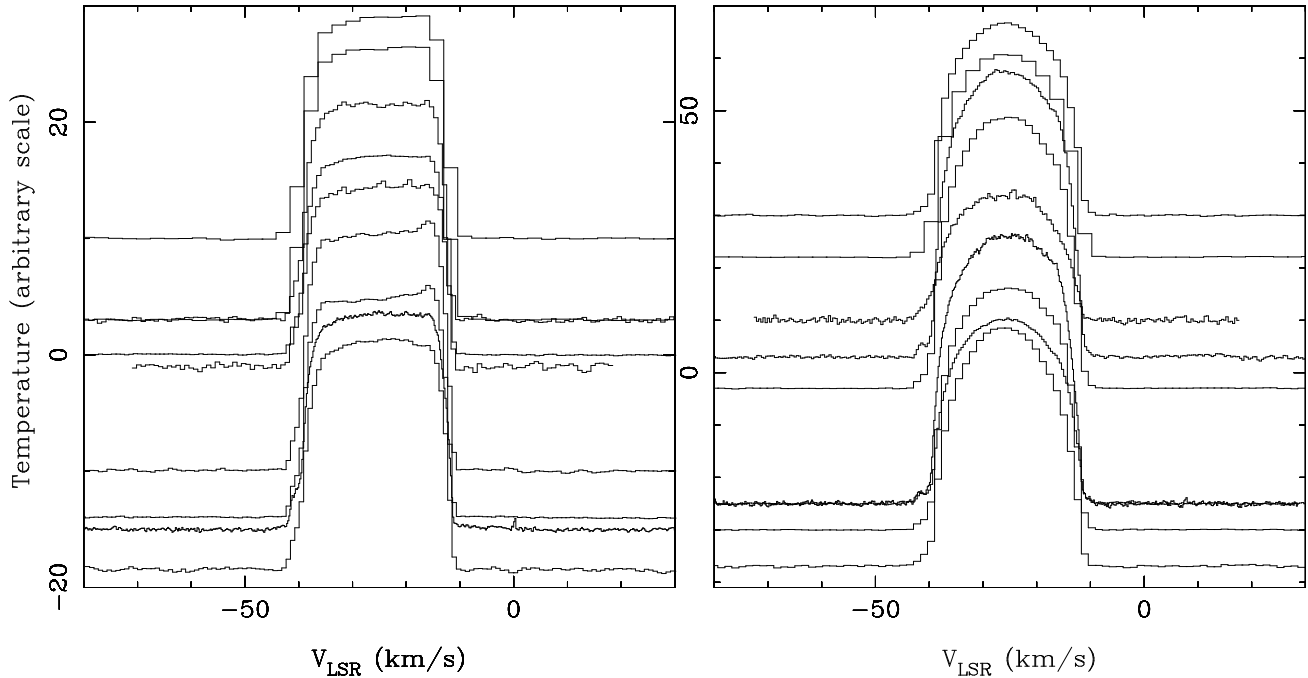,width=17.5cm,angle=-90}}
\caption[]{Time sequence of available IRAM J = 1-0 (left hand) and J = 2-1
(right hand) spectra. From top to bottom Sep. 1988, Aug. 1991, Dec. 1994,
Jan. 1995, May 1996, June 1996, Jan. 1997, April 1997, June 1998. 
Spectra are off-setted from each other.
}
\end{figure*}

\begin{table}
\caption[]{Data of spectra discussed}
\begin{tabular}{llllll} \hline
 Obs. date & J =  & $\Delta$v & \hspace{-0.3cm}r.m.s. & time   &\hspace{-0.3cm} phase \\
           &      &  (km/s)   & \hspace{-0.3cm} (K)   &  (min.) & $\phi$      \\ \hline
         September 1988       & 1-0 & 2.60 & \hspace{-0.3cm} 0.050 & 20 &\hspace{-0.3cm} 0.90 \\
                         & 2-1 & 1.30 & \hspace{-0.3cm} 0.088 & 20 &    \\
         7-10 August 1991  & 1-0 & 2.60 & \hspace{-0.3cm} 0.03  & 10 &\hspace{-0.3cm} 0.47 \\
                         & 2-1 & 2.60 & \hspace{-0.3cm} 0.06  & 10 &    \\
         23-29 December 1994 & 1-0 & 0.81 & \hspace{-0.3cm} 0.15  & 3  &\hspace{-0.3cm} 0.44 \\
                         & 2-1 & 0.41 & \hspace{-0.3cm} 0.18  & 3  &      \\
         25-28 January 1995 & 1-0 & 0.81 & \hspace{-0.3cm} 0.051 & 20 &\hspace{-0.3cm} 0.49 \\
                         & 2-1 & 1.30 & \hspace{-0.3cm} 0.055 & 20 &     \\
         28 May 1996     & 1-0 & 0.81 & \hspace{-0.3cm} 0.21  & 1  &\hspace{-0.3cm} 0.25 \\
                         & 2-1 & 0.41 & \hspace{-0.3cm} 0.39  & 3  & \\
         23 June 1996    & 2-1 & 0.20 & \hspace{-0.3cm} 0.29  & 8  &\hspace{-0.3cm} 0.28 \\    
         25 June 1996    & 1-0 & 0.26 & \hspace{-0.3cm} 0.14  & 4  &\hspace{-0.3cm} 0.28 \\
                         & 1-0 & 0.81 & \hspace{-0.3cm} 0.072 & 8  &     \\    
                         & 2-1 & 1.30 & \hspace{-0.3cm} 0.11  & 8  &  \\
         15-16 January 1997 & 1-0 & 0.26 & \hspace{-0.3cm} 0.17  & 16 &\hspace{-0.3cm} 0.59 \\
                         & 1-0 & 0.81 & \hspace{-0.3cm} 0.060 & 35 &     \\    
                         & 2-1 & 1.30 & \hspace{-0.3cm} 0.052 & 35 & \\
         3-4 April 1997   & 1-0 & 0.20 & \hspace{-0.3cm} 0.12  & 25 &\hspace{-0.3cm} 0.71 \\
                         & 2-1 & 0.41 & \hspace{-0.3cm} 0.13  & 25 &     \\    
                         & 1-0 & 3.38 & \hspace{-0.3cm} 0.20  & 25 &\hspace{-0.3cm} HCN \\
         3-4 June 1998   & 1-0 & 0.52 & \hspace{-0.3cm} 0.15  & 7 & \hspace{-0.3cm} 0.37 \\
                         & 1-0 & 0.81 & \hspace{-0.3cm} 0.14  & 7 &   \\    
                         & 2-1 & 1.30 & \hspace{-0.3cm} 0.16  & 14 &   \\    
\hline
\end{tabular}

\noindent
Listed are the date of the observing run, transition,
channel spacing in \ks, rms noise, integration time and phase in the
light curve.
\end{table}

\section{Comparison with other spectra}

IRC~+10~216 is an important millimetre line calibrator and therefore
many spectra taken with different telescopes over a long time
span are published in the literature; a compilation up to September
1992 can be found in Loup et al. (1993). Also one of us (MG), with
different collaborators, has observed this star many times.  As these
spectra are readily available we will concentrate on comparing these
spectra to the new data.  In addition, Dr. Truong-Bach (Observatoire
de Paris-Meudon) made his mapping data available described in
Truong-Bach et al. (1991), Dr. J. Kastner (MIT) who observed
IRC~+10~216 at IRAM in May 1996 made his high resolution J = 1-0 and
2-1 spectra available, and Dr. C. Thum (IRAM Grenoble) who was the
observer before us in June 1996 kindly made his high resolution J =
2-1 spectrum available.

\begin{figure}
\centerline{\psfig{figure=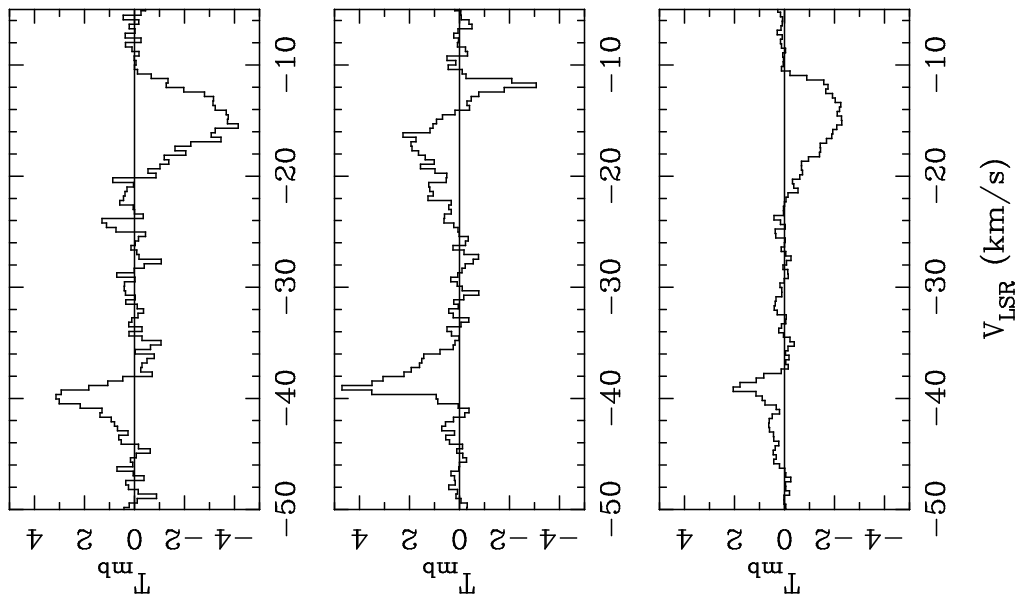,width=8.5cm,angle=-90}}
\caption[]{Difference between J = 2-1 spectra taken on 28 May 1996 and
December 1994 (top), between 23 June 1996 and December 1994 (middle) and
between April 1997 and December 1994 (bottom) at 0.4 \ks\ channel 
spacing. The December 1994 spectrum was scaled in each case 
before subtraction.
}
\end{figure}

Table 1 contains information on the spectra discussed in this paper
regarding observation dates, channel spacing, rms noise, integration
times and phase $\phi$ in the light curve, based on a period of 649
days and maximum light ($\phi$ = 0) at JD = 2447483 (Le Bertre 1992).

The CO J = 1-0 and 2-1 spectra are shown in Fig. 2, off-setted from
each other for clarity. We first discuss the J = 1-0 line. The
temperature at the line center of the IRAM spectra varies between
about 18 and 24 K. This is within the normally quoted absolute
calibration errors.

The shapes are all very similar with the exception that the June 1996
and January 1997 spectra clearly show the excess emission near the red
wing of the profile as discussed before. Other published IRAM 1-0
profiles [Bujarrabal et al. 1986 (observation date June 1985; $\phi
\sim$ 0.07) and Kahane et al. 1992 (observation date August 1988;
$\phi \sim$ 0.86)] are also all flat-topped with no evidence for
excess emission.

Regarding the J = 2-1 transition, there is no obvious evidence for
excess emission in the June 23 spectrum, nor in the lower resolution
June 25 spectrum, which is not shown here. To investigate the matter
further we present in Fig.~3 the difference spectra (May 1996 -
December 1994), (June 23, 1996 - December 1994) and (April 1997 -
December 1994), where we first adjusted the December 1994 spectrum to
the same absolute temperature as the spectrum from which it is
subtracted.  Evidence for excess emission is clearly seen between --23
and --15 \ks\ at a significant level in the June 1996 difference
spectrum, which is absent on the other two occasions. Furthermore,
there are features at --12 and --39 \ks\ in all three difference
spectra which indicate variations in the wings of the J = 2-1 line
profiles, possibly due to variations in the expansion velocity, or
small velocity shifts in the spectra due to optical depth effects.

We may conclude that the red excess emission was also present in the J
= 2-1 line in June 1996, but this is less clear and convincing than in
the J = 1-0 spectra. The reason is that the velocity interval where
the excess occurs, lies on the steep wing of the 2-1 profile but on
the essentially flat-topped 1-0 profile, and therefore is easier to
detect. In a way we are fortunate that the combination of source size
of the CO (1-0) emission and the beam size of the IRAM telescope at
that frequency result in a flat-topped profile.

\section{The mapping data}

In April 1997 we mapped IRC~+10~216 out to a distance of about
110\arcsec\ in the CO (1-0), CO (2-1) and HCN (1-0) lines.  The only
other published IRAM CO map of this object is by Truong-Bach et
al. (1991) who mapped the star out to 54\arcsec\ at 1 MHz spectral
resolution. Our CO map was obtained at 78 KHz and 312 KHz resolution.

The on-source spectra are shown in Figs.~4 (CO J = 1-0), 5 (CO J =
2-1) and 6 (HCN 1-0). Some of the spectra at large radial off-sets are
smoothened. The CO spectra are normalised so they appear equally strong.
The absolute scale can be estimated from the zero intensity level,
which is --0.3 K in all plots.

The off-set spectra in CO are remarkable, in the sense that the
profiles are not smooth but show distinct components. This has never
been seen before in this, or any other, AGB star. The HCN map does not
show this behaviour. This is at least partly due to the much lower
velocity resolution, which is probably also the reason why Truong-Bach
et al. (1991) did not see these components in their CO spectra (see below).

One can identify components between approximately --18 and --14 \ks\
(strikingly similar in velocity to the excess emission seen in June
1996 and January 1997), --40 and --34 \ks, --34 and --22 \ks. The exact
velocities seem to change slightly with position. The relative
strength of these components changes dramatically with position. On
the other hand, there is good correspondence between the components in
the 1-0 and 2-1 profiles at a given off-set position.

We have made contour plots of the integrated intensity for different
velocity intervals. Because the step size in the map is larger than the
size of the beam this analysis is inconclusive as to tell whether
different velocity bins peak at different positions. However, the peak
of the emission is not at the central position but slightly to the
north (see below as well).

Our data and that by Truong-Bach et al. (1991) are taken 8.5 years
apart with the same telescope and it is interesting to compare the two
data sets. In their Fig.~3 they plot the 1-0 and 2-1 profiles at 0, 6,
12, 24, 36 and 54\arcsec\ off-set, averaged over the north, east,
south west position. We have done the same for our spectra at 10, 30
and 36\arcsec\ off-set, to compare to their spectra at 12, 36 and
54\arcsec\ off-set. We have resampled our data to their 1 MHz
resolution. The two data sets are compared in Fig.~7. Represented in
this way, it is clear that the two datasets are indistinguishable. This
implies that these phenomena {\it could} have been present at that
time as well.

From Fig.~6 it is clear that the peak of the HCN integrated intensity
is not at the central position. From a contour plot of the integrated
intensity we estimate it to be at (+5\arcsec, +10\arcsec). The
emission is elongated as well with a position angle of about
45\degr. This is contrary to the previous interferometric observations
of Bieging et al. (1984) and Dayal et al. (1995), where the HCN
emission is centered on the central position, and approximately spherical.
A pointing error seems very unlikely, as the pointing was checked
regularly on the same nearby pointing source, and the rms errors in
the pointing model were of order 3\arcsec\ only.

\begin{figure*}
\centerline{\psfig{figure=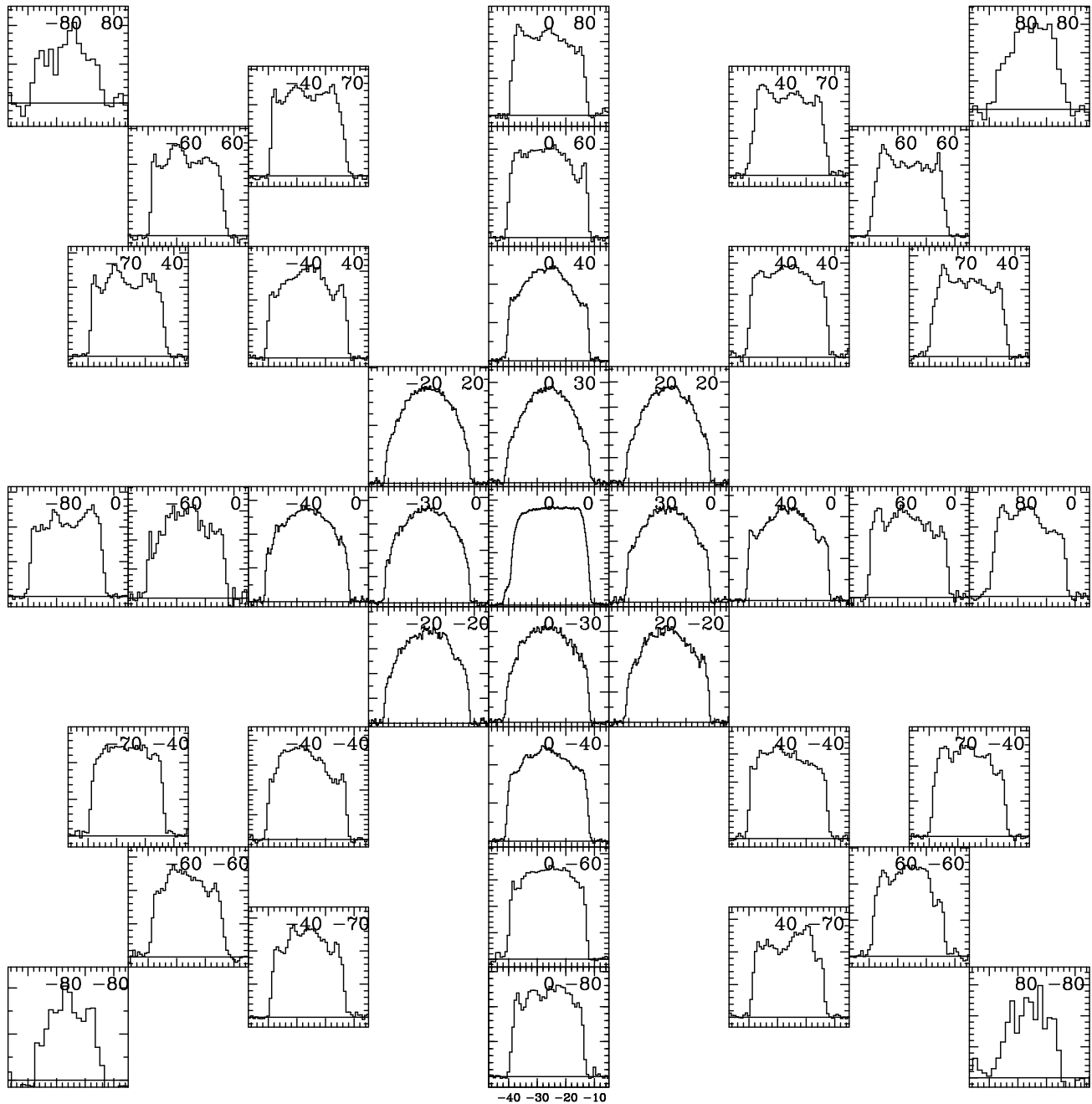,width=17.5cm,angle=-90}}
\caption[]{CO 1-0 map. Offsets are indicated in each panel. All
spectra are normalised, to better illustrate the different
components. The lower temperature limit plotted is --0.3 K in all
cases. The velocity range shown is between --47 and --5 \ks. North is
up, East to the right.
}
\end{figure*}

\begin{figure*}
\centerline{\psfig{figure=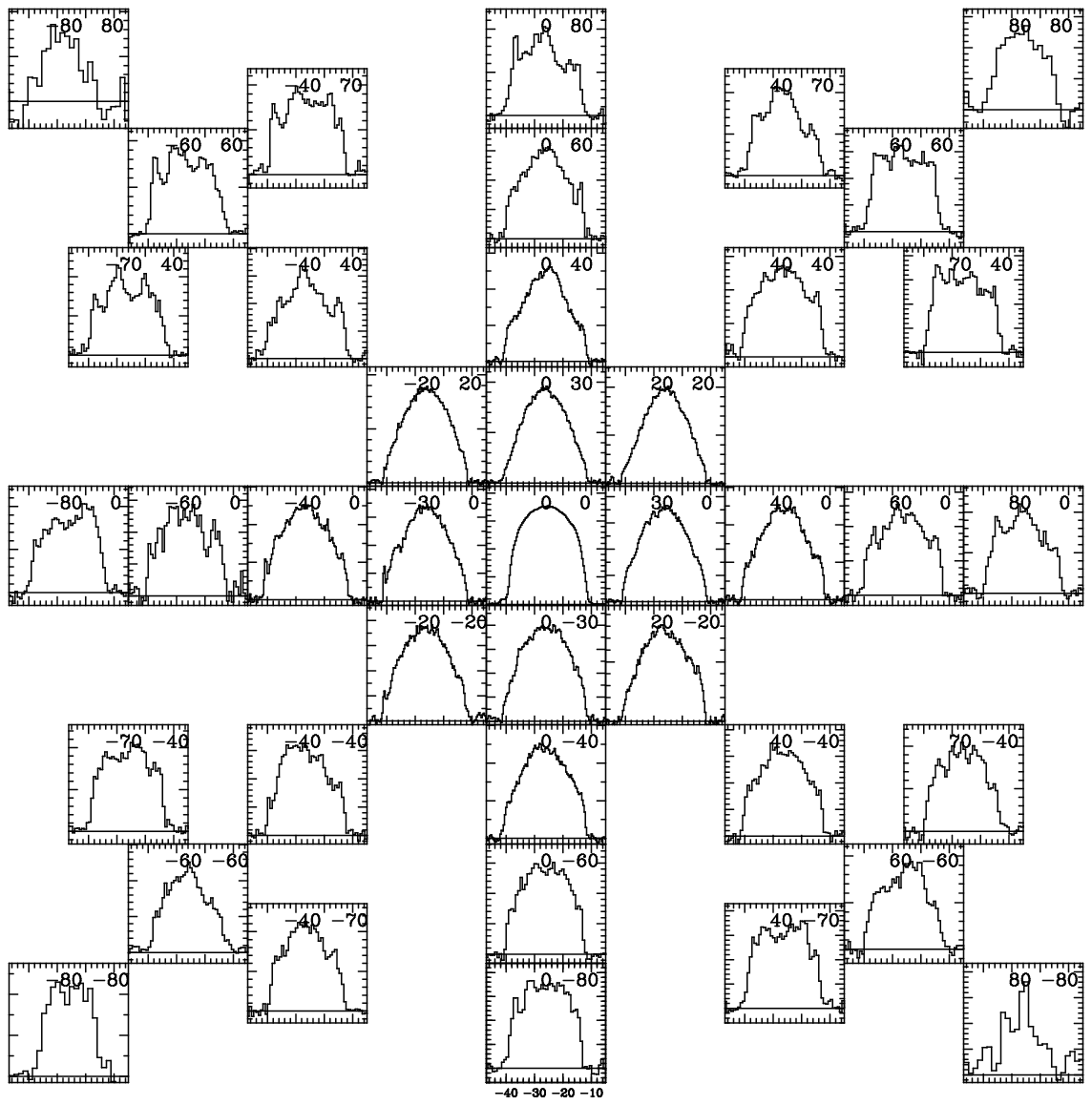,width=17.5cm,angle=-90}}
\caption[]{CO 2-1 map. As Figure~4.
}
\end{figure*}

\section{Discussion}

We have discovered two -- probably related -- new and interesting features
in the prototype infrared carbon star IRC~+10~216: a transient
emission feature in the red wing of the CO 1-0 profile (also present but
less clear in the 2-1 profile), and different emission components in
both 1-0 and 2-1 profiles throughout the envelope. \\

Although these observations speak for themselves, the interpretation
of these phenomena is less straightforward. One firm conclusion is
that the circumstellar CO shell is not spherically symmetric. \\

Asymmetries close to the star on a scale of 0.1-0.2\arcsec\ were known
previously from infrared observations (see Weigelt et al. 1998, Haniff
\& Buscher 1998 for the latest on this) but we now show that there
exist asymmetries throughout the envelope. \\

\begin{figure}
\centerline{\psfig{figure=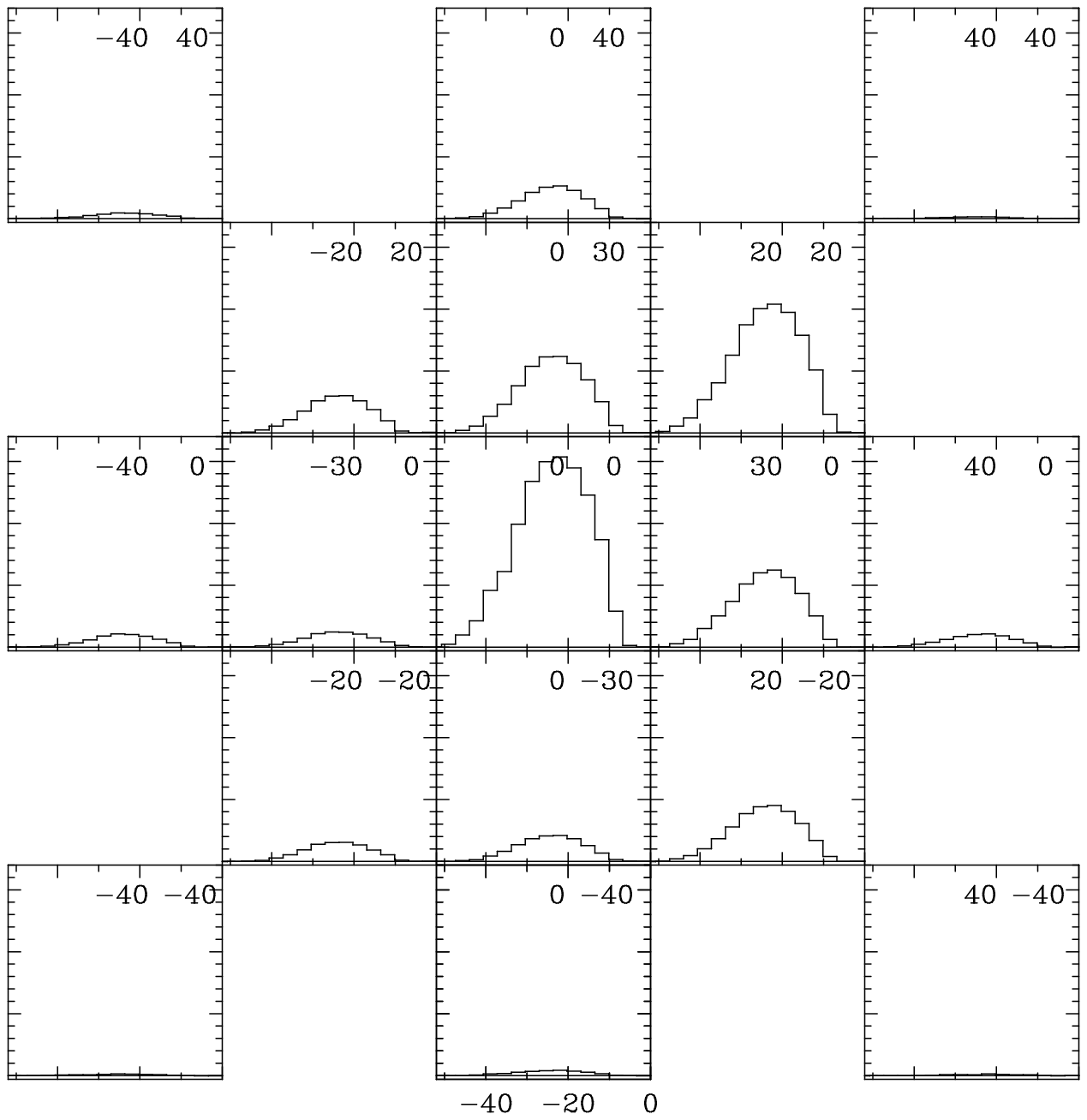,width=8.5cm,angle=-90}}
\caption[]{HCN 1-0 map. The velocity range plotted is between --52 and --0
\ks, while the intensity is between --0.3 and 17 K. North is
up, East to the right.
}
\end{figure}

Let us first point out that the original discovery of the excess
emission in the on-source spectrum, as well as the later notion of
features in the off-set spectra could only have been discovered thanks
to the high velocity resolution (of order 100 KHz or better). As we
demonstrated by comparing our resampled data to the map of Truong-Bach
et al. (1991), a resolution of 1 MHz is insufficient to detect these
features.\\

A second point to note is that an observer who uses IRC~+10~216 as a
calibration object, but has little or no experience or interest in the
CO spectra of carbon stars as such, would perhaps not note or care
that a CO 1-0 spectrum showed some small changes with respect to the
calibration spectrum. In order words, it might well be that this
feature has been observed before, but went unnoticed. It might
therefore be worthwhile to look at archival data (in particular high
resolution CO 1-0 spectra) taken with various telescopes. Published
higher-J CO data (e.g. compilation in Groenewegen et al. 1998) do not
seem to show this feature. \\

One question that might be addressed if such features where to be found
in archival data, is the recurrence time scale of these phenomenon. Was
its presence in the on-source spectra, that started between May and
June 1996, and ended between January 1997 and April 1997, unique ?
This is a time span of between 7 and 11 months, compared to the
pulsation period of about 21 months. Could there be a connection ?
On the other hand, if this phenomenon is connected to pulsation and
would occur every cycle for 30-50\% of the time, then it seems
improbable that it has never been seen before, even considering the biases
discussed above against observing such a phenomenon. \\

If something came in and out the line of sight, this would imply it
has traversed roughly the full width of the beam (21\arcsec) in 7-11
months. At a distance of 135 pc, this corresponds to a traverse motion of
15-23 $\times$ 10$^3$ \ks, improbably high. \\

One of the first things that comes to mind to explain the mapping data
is a geometrical one, for example a disk, or a ``shell with evacuated
holes'' (Sloan \& Egan 1995), as for example visualised in Dyck et
al. (1987; their Fig.~7 which was based to explain various
observations on a scale of a few arc seconds, much smaller than the
scale we are considering now).  However, in such a schematic picture
(assuming a disk with an uniform outflow and no angle dependence of
the emission) one expects point symmetry of the excess emission around
the central position of the red and blue-shifted emission. This is
clearly not the case. \\

Knapp et al. (1998) present evidence of two winds with different
velocities in several AGB stars. They observed J = 2-1 and 3-2 spectra
at high velocity resolution. In particular the J = 3-2 spectrum of R
Leo is qualitatively similar to our on-source J = 1-0 spectrum of IRC~+10~216,
having red excess emission. This is in contrast to their other
examples (e.g. IRC~+50~049, X Her [also see Kahane \& Jura 1996], EP
Aqr), where the two winds can be seen symmetrically w.r.t. the stellar
velocity at both blue and red shifted velocities. 

So, neither of these simple explanations seem to apply to our
observations. Although the explanation of the present observations
probably lies in a geometrical one, we can not constrain this geometry
at present. Additional mapping data in the inner part at a higher
spatial resolution and archival (on-source) data will be of help. \\

\begin{figure*}
\centerline{\psfig{figure=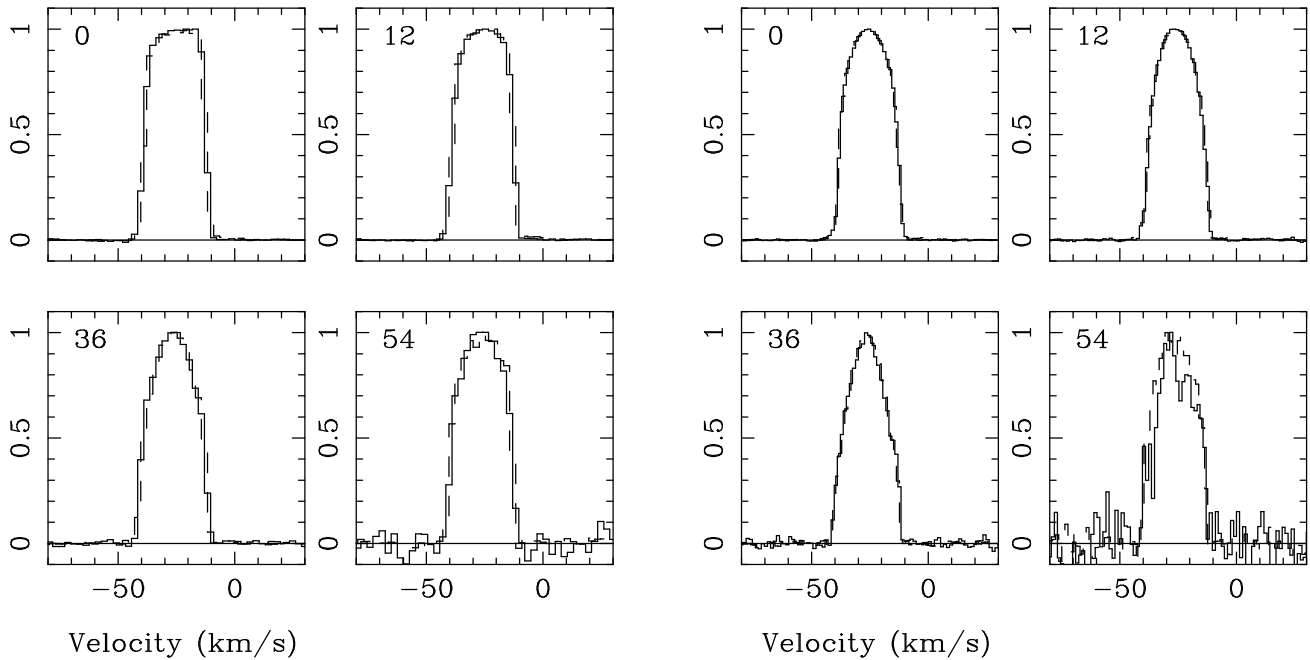,width=17.5cm,angle=-90}}
\caption[]{A comparison between our data (dashed line) and that of
Truong-Bach et al. (1991; solid line), averaged over N,E,S,W positions
at the (approximate) off-sets indicated in the panels for the J = 1-0
(left hand side) and J = 2-1 (right hand side). Our data has been
smoothened to their spectral resolution, and both datasets are
normalised. The datasets are taken 8.5 years apart, and represented in this
way, are indistinguishable. }
\end{figure*}

We will now focus on the transient behaviour of the red excess
emission in the on-source spectrum. To pursue to idea of a clump
ejected by the star, consider the following. Solving the radiative
transfer equation, and assuming optically thin emission, and an
excitation temperature much larger than the cosmic background
temperature, one can derive the following expression between column
density and integrated emission (e.g. Johansson et al. 1984):
\begin{equation}
\frac{N_l}{g_l} = \frac{1.67 \times 10^{14}}{\nu \, {\mu}^2 \, S}
\, \int{T_{\rm mb} \, dv},
\end{equation}
with $\nu$ in GHz, $\mu$ the permanent dipole moment (in Debye), the
integrated intensity in \kks, $S$ the line strength which is equal to
the J-number of the upper level in the case of CO and $N_l$ and $g_l$
the column density and statistical weight of the lower level.
Furthermore, under the assumption of LTE, one has
\begin{equation}
\frac{N_l}{g_l} = \frac{N_{\rm tot}}{Q(T_{\rm ex})} \, \exp(-E_l/kT_{\rm ex}),
\end{equation}
with $Q$ the partition function, and $N_{\rm tot}$ the total column
density.  For excitation temperatures for the CO(1-0) line between 15
and 50 K, assuming an abundance ratio of CO-to-H$_2$ of 1.1 $\times
10^{-3}$ (Groenewegen et al. 1998), one then finds within a factor of 2 that
\begin{equation}
N_{\rm H_2} = 1.1 \times 10^{18} \int{T_{\rm mb}(1-0) \, dv} 
\hspace{15mm}{\rm cm^{-2}}.
\end{equation}
If the tempereture is higher than 10-50 K, the column density goes up;
for example, for $T_{\rm ex} = 80$ K the factor in front would be $3.2
\times 10^{18}$.

The largest blob identified in the Weigelt et al. (1998) paper has a
radius of about 30 milli-arcseconds which, at 135 pc, corresponds to a
linear size of $5 \times 10^{13}$ cm. This is indeed somewhat larger
than the typical blob sizes considered (Olofsson 1994, Olofsson et
al. 1996).  Recalling that in June 1996 the excess emission had an
CO(1-0) integrated intensity of 2.4 \kks, and combining all this, one
finds that the density in such a single large blob is $5.2 \times
10^4$ H$_2$/cm$^3$.  Since the blob size (60 mas diameter) is much
smaller than the beam size (22\arcsec\ FWHM), and Eq. (1) assumes that
the source size is comparable to the beam size, this estimate is likely
to be too low by a factor $(22000/60)^2 \approx 1.3 \times 10^5$. This
then would lead to a density estimate of $\sim 7 \times 10^{9}$
cm$^{-3}$ within a factor of a few. This is similar to the estimate by
Olofsson (1994) of $10^{10}$ cm$^{-3}$ based on SiO maser spots. The mass is
this blob would be $8 \times 10^{-6}$ \msol\ allowing for 10\% helium.

A distance of $10^{14}$ cm is crossed in 2.2 years at the expansion
velocity of 14.5 \ks. With an underlying average mass loss rate of
about $1.8 \times 10^{-5}$ \msolyr (e.g. Groenewegen et al. 1998)
there would be enough mass available to form such a single big blob.
This ``quantitative estimate'' shows that is it possible that a single
large blob (possibly only rarely formed) is responsible for the excess
emission. If one entertains the possibility of blobs further, and
notices that an integrated emission which is about a factor of 5
smaller than the 2.4 \kks\ in June 1996 would not have been detected,
one can conclude that a similar phenomenon but now a mass ejection in
the form of 100 clumps of size 10$^{13}$ cm in random directions
probably would have little observational consequences. The same goes
for even more clumps of smaller size.

The excess emission at the red wing implies that the blob is moving
away from us. Its sudden appearance could then be related to the fact
that the blob was previously occulted by the central star. The fact
that the excess emission disappeared again could either mean that the
blob moved again behind the star or that the column density decreased. 
The former possibility is unlikely as it requires a non-radial motion
and implies velocity shifts which are not readily seen in the spectra
obtained at later dates. The latter possibility seems physically more
attractive. The width of about 4 \ks\ in the excess emission in June
1996 already implies that there is some differential expansion or
turbulent motion. Over a period of 10 months and with a differential
velocity of 4 \ks\ the blob would be of size $6 \times 10^{13}$ cm, and
hence the column density would have decreased by about 45\%. This
would be too little to explain the observations which might indicate
higher internal motions, or that the big blob broke up into smaller
blobs following slightly different trajectories. \\

\acknowledgements{The authors would like to thank Clemens Thum (IRAM
Grenoble) for providing his June 23, 1996 J = 2-1 spectrum and helpful
comments, Joel Kastner (MIT) for providing his 28 May 1996 J = 1-0 and
2-1 spectra, Truong-Bach (DEMIRM, Observatoire de Paris-Meudon) for
his 1991 data and Wolfgang Wild (IRAM Grenada) for helpful comments. 
The referee, T. Le Bertre, is thanked for comments and suggestions 
that improved the paper.
} \\

{}

\end{document}